\documentclass[twoside,reqno,11pt]{amsart}
\newtheorem{thm}{Theorem}[section]

\theoremstyle{definition}
\newtheorem{defn}[thm]{Definition}
\theoremstyle{remark}
\newtheorem{rem}[thm]{Remark}
\newtheorem*{ex}{Example}

\numberwithin{equation}{section}

\newcommand{\symb}[1]{{\tt #1}}

\renewcommand{\vec}[1]{\boldsymbol{\mathrm{#1}}}

\newcommand{\mvectdiff}[1]{ \nabla_#1 }

\newcommand {\grpart}[1]{\ensuremath{\left\langle #1 \right\rangle  }}

\usepackage{amsfonts, amsmath, amssymb,latexsym}
\usepackage[english]{babel}
\usepackage{url}
\usepackage{graphicx}
\usepackage[usenames,dvipsnames]{pstricks}
\usepackage{epsfig}
\usepackage{color}
\usepackage{listings}
\usepackage{color}
\usepackage{comment}
\usepackage{framed}
\usepackage{float}

\definecolor{labelcolor}{RGB}{100,0,0}
\definecolor{outputcolor}{RGB}{0,0,100}
\colorlet{shadecolor}{yellow!30}
\definecolor{dkgreen}{rgb}{0,0.6,0}
\definecolor{gray}{rgb}{0.5,0.5,0.5}
\definecolor{mauve}{rgb}{0.20 , 0.40, 1.0}

\floatstyle{plaintop}
\newfloat{listing}{thp}{lop}
\floatname{listing}{Listing}

\makeatletter
\renewcommand{\verbatim@font}{\footnotesize\ttfamily}
\makeatother

\begin{document}

\title[Clifford and Tensor Algebras in Maxima]{Sparse Representations of Clifford and Tensor algebras in Maxima}

\author{Dimiter Prodanov\textsuperscript{1}}

\address{
	Department of Environment, Health and Safety,
	Neuroscience Research Flanders,
	IMEC vzw, Leuven, Belgium
}
\email{dimiterpp@gmail.com; \, dimiter.prodanov@imec.be }

\author{Viktor T. Toth \textsuperscript{2}}
\address{Center for Research on Integrated Sensors Platforms
	Carleton University
    Ottawa, Ontario, Canada
}
\email{vttoth@vttoth.com}



\newcommand{\Addresses}{{
		\bigskip
		\footnotesize
		
	 \textsuperscript{1} Department of Environment, Health and Safety,
	 Neuroscience Research Flanders,
	 IMEC vzw, Leuven, Belgium
		
	\medskip
		
	\textsuperscript{2} Center for Research on Integrated Sensors Platforms
	Carleton University
	Ottawa, Ontario, Canada		
	}}
	
\begin{abstract}
Clifford algebras have broad applications in science and engineering. The use of Clifford algebras can be further promoted in these fields by availability of computational tools that automate tedious routine calculations.
We offer an extensive demonstration of the applications of Clifford algebras in electromagnetism using the geometric algebra $\mathbb{G}^3 \equiv C\ell_{3,0}$ as a computational model in the Maxima computer algebra  system.
We compare the geometric algebra-based approach with conventional symbolic tensor calculations supported by Maxima, based on the \symb{itensor} package.
The Clifford algebra functionality of Maxima is distributed as two new packages  called  \symb{clifford} - for basic  simplification of Clifford products, outer products, scalar products and inverses; and \symb{cliffordan} - for applications of geometric calculus.
\end{abstract}

\maketitle

\Addresses

\section{Introduction}
\label{sec:intro}

While Computer Algebra Systems (CAS) cannot replace mathematical intuition, they can facilitate routine calculations and teaching.
This paper focuses on applications of two packages implementing abstract algebras in the popular open-source {\em Maxima} CAS:
the \symb{itensor} package implementing indicial tensor manipulation and a new package for Clifford algebra called \symb{clifford}, along with its companion package \symb{cliffordan}.
While an extensive overview of the Maxima system is not our objective, we offer some details that demonstrate the major advantages of Maxima over purely numerical systems, such as Matlab (Mathworks, Natick, MA, USA).

\subsection{System level functionality}
\label{sec:maxintro}

Maxima is derived from one of the first ever computer algebra systems, MACSYMA (Project MAC's SYmbolic MAnipulation System),  developed by the Mathlab group of the MIT Laboratory for Computer Science during the years 1969-1972.
Maxima is written entirely in Lisp and is distributed under open source license\footnote{Maxima is distributed under GNU General Public License and developed and supported by a group of volunteers}.
Maxima has its own programming language, which is particularly well-suited for handling symbolical mathematical expressions or mixed numerical-symbolical expressions.
In addition, the system supports 64-bit precision floating point and arbitrary precision arithmetics.
Maxima programs can be automatically translated and compiled to Lisp within the program environment itself.
Third-party Lisp programs can be also loaded and accessed from within the system.
The system also offers the possibility of running batch unit tests.

Maxima supports several primitive data types \cite{M2015}:
\textit{numbers} (rational, float and arbitrary precision);
\textit{strings} and \textit{symbols}.
In addition there are compound data types, such as \textit{lists},
 \textit{arrays}, \textit{matrices} and \textit{structs}.
There are also special symbolic constants, such as the Boolean constants \symb{true} and \symb{false} or the complex imaginary unit \symb{\%i}.

Several types of operators can be defined in Maxima.
An operator is a defined symbol that may be  \textit{unary} \symb{prefix}, \textit{unary} \symb{postfix}, \textit{binary} \symb{infix}, \textit{n-ary} \symb{matchfix}, or \symb{nofix} types.
For example, the inner and outer products defined in \symb{clifford} are of the \symb{binary infix} type.
The scripting language allows for defining new operators with specified precedence or redefining the precedence of existing operators.
Maxima distinguishes between two forms of applications of operators -- forms which are \symb{noun}s and forms which are \symb{verb}s.
The difference is that the \symb{verb} form of an operator evaluates its arguments and produces an output result, while the \symb{noun} form appears as an inert symbol in an expression, without being executed.
A \symb{verb} form can be mutated into a \symb{noun} form and vice-versa.
This allows for context-dependent evaluation, which is especially suited for symbolic processing.

\subsection{Expression representation and transformation in Maxima}
\label{sec:expr}

\begin{figure}[ht]
\includegraphics[width=0.4\linewidth]{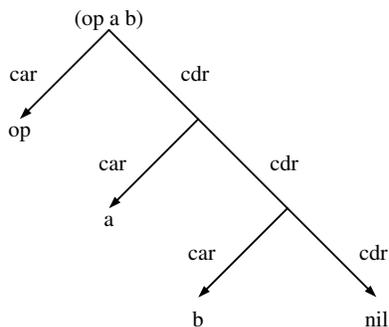}
\caption{\label{fig:lisp}Expression representation in Lisp. A general expression can be represented as a list with the first element being the operator  \symb{op} and the rest of the elements representing operator arguments. For instance, the expression \symb{op(a,b)} will be represented by the list \symb{(op a b)}, which is the ordered pair of the atom \symb{op}, and another list, \symb{(a b)}, which, in turn, is represented by another ordered pair.}
\end{figure}

\begin{figure}[ht]
\includegraphics[width=0.6\linewidth]{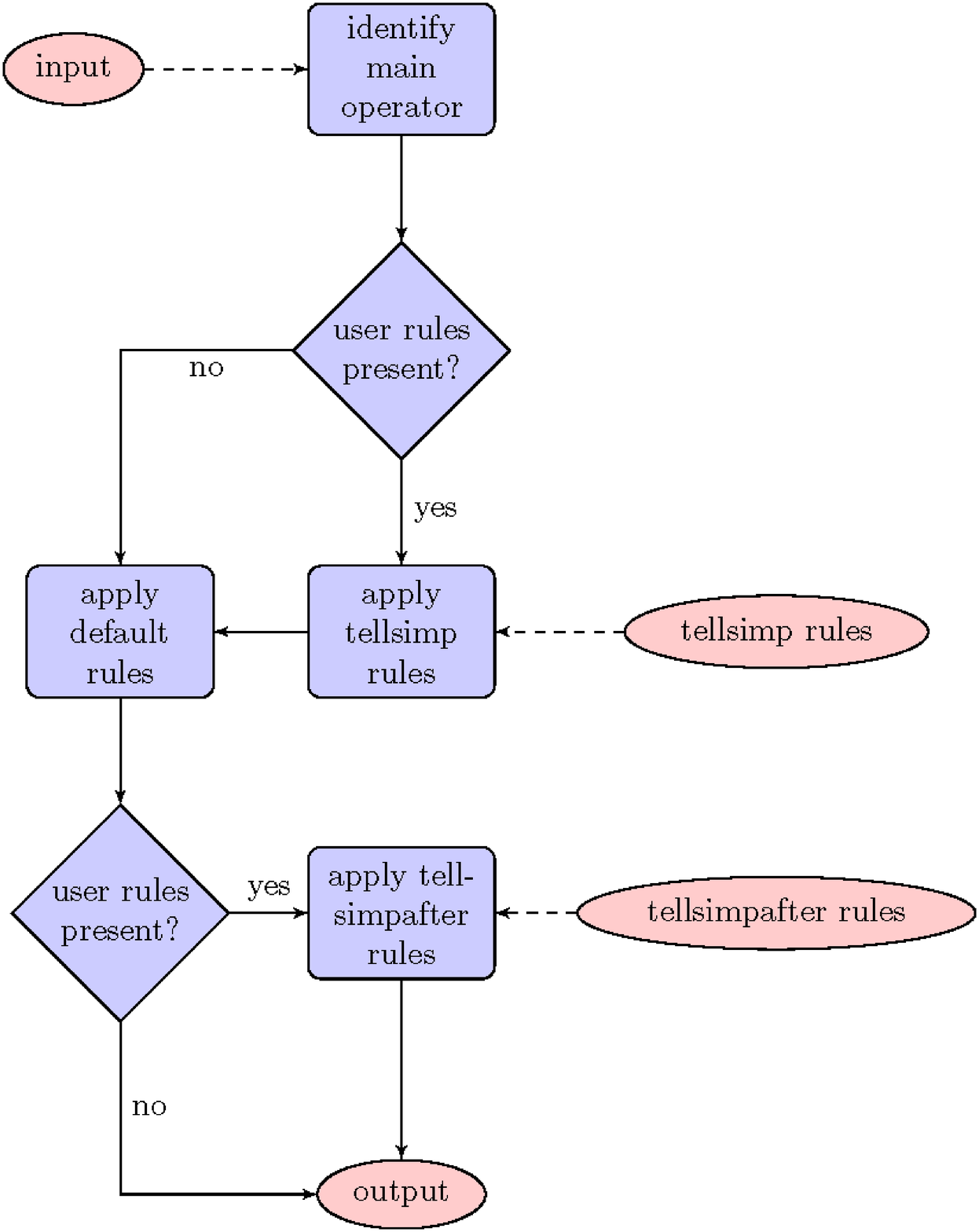}	
\caption{Expression simplification in Maxima.}
\label{fig:flowchart}
\end{figure}

\begin{listing}
\caption{\label{lst:definitions}Dot products and exponent simplification rules in   \symb{clifford}.}
{\color{labelcolor}\begin{verbatim}
if get('clifford,'version)=false then (
	tellsimp(aa[kk].aa[kk], signature[kk] ),
	tellsimpafter(aa[kk].aa[mm], dotsimp2(aa[kk].aa[mm])),
	tellsimpafter(bb.ee.cc, dotsimpc(bb.ee.cc)),
	tellsimpafter(aa[kk]^nn, powsimp(aa[kk]^nn)),
	tellsimpafter(aa[kk]^^nn, powsimp(aa[kk]^^nn))
);
\end{verbatim}}
\end{listing}

The manner in which Maxima represents expressions, function calls and index expressions using the Lisp language is particularly relevant for the \symb{itensor} and \symb{clifford} packages. In the underlying Lisp representation a Maxima \textit{expression} is a tree containing sequences of operators, numbers and symbols. Every Maxima expression is simultaneously also a \textit{lambda} construct and its value is the value of the last assigned member. This is a design feature inherited from Lisp. Maxima expressions are represented by underlying Lisp constructs. The core concept of the Lisp language is  the idea of a  \textit{list} representation of the language constructs. Or, to be more precise, the idea of an ordered pair, the first element of which is the head (or {\tt car}), the second element the tail (or \symb{cdr}) of the list  (see. Fig. \ref{fig:lisp}). List elements are themselves either \textit{lists} or   \textit{atoms}: e.g., a number, a symbol, or the empty list (\symb{nil}). This representation enables the possibility to define transformation rules. In such way a part of an expression can be matched against a pattern and transformed to another expression.

A very powerful feature of the system is the ability to define custom transformation rules. Various transformation rules can be associated with any given operator in Maxima. Maxima has an advanced pattern matching mechanism, which supports nesting of operators and simplification (see Fig. \ref{fig:flowchart}). The simplifier subroutine operates by descending the tree of an expression until it gets to atoms, and then simplifies the smallest pieces and backs out.

User-defined rules can be added to the built-in simplifier using one of two  commands: \symb{tellsimp} or \symb{tellsimpafter}. Rules in both sets are identified by the main operator. Rules specified using \symb{tellsimp} are applied  before the built-in simplification, while \symb{tellsimpafter} rules are applied after the built-in simplification.
The augmented simplification is then treated as built-in, so subsequent tellsimp rules are applied before those defined previously.
An example is given in Listing~\ref{lst:definitions} used in the implementation of \symb{clifford}.

This rule-based approach to computer algebra is particularly suitable to implement ``sparse'' representations of algebraic objects. For instance, a tensor object may be defined only in terms of its indices and the transformation rules that it obeys, without ever assigning values to tensor components. This is how tensors are implemented in the \symb{itensor} package of Maxima. Similarly, a multivector can be defined symbolically as a linear combination of basis vectors, without ever assigning those basis vectors any specific values or representing them by matrices.
This is the approach followed by the built-in \symb{atensor} package and the \symb{clifford} package that is described in the present paper.

\section{Tensor algebra representations in Maxima}
\label{sec:itensor}

Tensors are abstract objects that describe multilinear relations between geometric objects.
A coordinate-independent way to define tensors is by multilinear maps.
In this approach, a tensor $T$ of type $(p, q)$ is defined as a map that is linear in each of its arguments, by the equation
\begin{equation}
  T: \underbrace{V^\star \otimes \ldots V^\star}_{p } \otimes \underbrace{V \otimes \ldots V}_{q}  \rightarrow \mathbb{K},
\end{equation}
where $\mathbb{K}$ is a field, $V$ is a (finite-dimensional) vector space over $\mathbb{K}$, while $V^\star$ is the corresponding dual space of co-vectors.

Maxima has two major packages for tensor manipulation. The \symb{ctensor} package implements component representation of tensors, with functionality that is especially tailored for applications in general relativity. The \symb{itensor} package treats tensors as opaque objects, manipulated via their indices. As such, \symb{itensor} is especially suited for computations where general covariance is maintained, including general relativity. The package has built-in support for the metric and curvature tensors, for covariant differentiation, for the Kronecker delta, and for utilizing the symmetry properties of tensors for algebraic simplification. All this is accomplished without the need to define tensor components.

The \symb{itensor} package also has facilities for tensor calculus, including functional differentiation with respect to tensor quantities. These facilities make it possible to use \symb{itensor} to investigate field theories that are formulated using a Lagrangian density functional. By way of a demonstrative example, let us consider the Lagrangian density for the Maxwell field (hereafter we assume that indices run from 1 through the number of dimensions $D$, and we use the Einstein summation convention, $x_\mu y^\mu=\sum_{\mu=1}^D x_\mu y^\mu$):
\begin{equation}\label{eq:tenslag}
  \mathcal{ L}_{\rm EM} = -\frac{1}{4}F_{\kappa\lambda}F^{\kappa\lambda}+j_\kappa A^\kappa \, ,
\end{equation}
where $F_{\kappa\lambda}=\partial_\kappa A_\lambda-\partial_\lambda A_\kappa$ is the electromagnetic field tensor corresponding to the 4-potential $A_\mu$.

The Euler-Lagrange equation that corresponds to this Lagrangian is written as

\begin{equation}
\frac{\partial(\sqrt{-g}\mathcal{L}_\mathrm{ EM})}{\partial A_\mu}-\nabla_\nu\frac{\partial(\sqrt{-g}\mathcal{L}_\mathrm{EM})}{\partial(\nabla_\nu A_\mu)}=0,
\label{eq:MaxLag}
\end{equation}
where $\nabla_\nu$ is the covariant derivative with respect to the coordinate $x^\mu$.

For the sake of simplicity of discussion, we restrict the presentation to rectilinear coordinates of special relativity, such that the square root of metric determinant $\sqrt{-g}$ can be assumed to be constant and removed from consideration, and covariant derivatives can be replaced with ordinary partial derivatives with respect to the coordinates, i.e., $\partial_\mu=\partial/\partial x^\mu$. Under these circumstances, the Euler-Lagrange equation becomes

\begin{equation}
\frac{\partial \mathcal{L}_\mathrm{EM}}{\partial A_\mu}-\partial_\nu\frac{\partial\mathcal{ L}_{\rm EM}}{\partial(\partial_\nu A_\mu)}=0.
\end{equation}

The first term in this equation is trivial: $\partial\mathcal{L}_\mathrm{ EM}/\partial A_\mu=j^\mu$, as follows directly from the definition of $\mathcal{ L}_\mathrm{ EM}$. As to the second term, we first note that

\[
F_{\kappa\lambda}F^{\kappa\lambda}=F_{\kappa\lambda}g^{\kappa\alpha}g^{\lambda\beta}F_{\alpha\beta},
\]
and that
\[
\frac{\partial F_{\kappa\lambda}}{\partial(\partial_\nu A_\mu)}=\delta^\nu_\kappa\delta^\mu_\lambda-\delta^\nu_\lambda\delta^\mu_\kappa,
\]
thus
\begin{equation}
\frac{\partial}{\partial(\partial_\nu A_\mu)}F_{\kappa\lambda}F^{\kappa\lambda}=(\delta^\nu_\kappa\delta^\mu_\lambda-\delta^\nu_\lambda\delta^\mu_\kappa)F^{\kappa\lambda}+
 F^{\alpha\beta}(\delta^\nu_\alpha\delta^\mu_\beta-\delta^\nu_\beta\delta^\mu_\alpha)=4F^{\nu\mu}=-4F^{\mu\nu}.
\end{equation}
Therefore
\begin{equation}
\partial_\nu\frac{\partial\mathcal{L}_\mathrm{ EM}}{\partial(\partial_\nu A_\mu)}=-\frac{1}{4}\partial_\nu\frac{\partial}{\partial(\partial_\nu A_\mu)}F_{\kappa\lambda}F^{\kappa\lambda}=\partial_\nu F^{\mu\nu}=F^{\mu\nu}_{,\nu},
\end{equation}
where we used the common shorthand notation $\partial_\mu A=A_{,\mu}$ for ordinary partial coordinate derivatives. Finally, we can now put everything together and obtain the explicit form of the Euler-Lagrange equation, which amounts to the definition of the 4-current:
\begin{equation}\label{eq:MaxTens}
 j^\mu+F^{\mu\nu}_{,\nu}=0.
\end{equation}

\begin{ex}
In the following Maxima code snippets, input expressions in Maxima are sequentially labeled by \symb{\%i{\em N}}, while the output may be labeled by \symb{\%o{\em N}}, where {\em N} is a number, starting with 1, that is incremented by 1 for each new input line. Commands are grouped together between parentheses. Multiple commands can be entered by separating them with a comma. The semicolon or the dollar sign terminate a command line; the latter can be used if the suppression of output is desired (e.g., to avoid cluttering the terminal display with very large intermediate expressions.) For clarity of presentation the output labels are suppressed and the expressions are colored. Expressions are only minimally edited for improvement of readability.

To carry out the derivation (\ref{eq:MaxLag})--(\ref{eq:MaxTens}) using \symb{itensor} in Maxima, after loading the package and configuring the metric (see Listing~\ref{lst:tensor} in the Appendix), we define the tensor $F_{\mu\nu}$ in terms of its components and construct $\mathcal{ L}_\mathrm{EM}$\footnote{The {\tt ishow} command is a built-in ``pretty print'' feature of {\tt itensor} to display tensor quantities}:
{\color{labelcolor}\begin{verbatim}
 (%i3) components(F([m,n],[]),A([n],[],m)-A([m],[],n))$
 (%i4) L:ishow(-1/4*F([k,l])*F([a,b],[])*g([],[k,a])*g([],[l,b])
       +j([k],[])*A([l],[])*g([],[k,l]))$
\end{verbatim}}\vskip -1em
\[\color{outputcolor}
{{g}^{\mathit{k l}}}\, {{j}_{k}}\, {{A}_{l}}-\frac{{{g}^{\mathit{k a}}}\, {{g}^{\mathit{l b}}}\, \left( {{A}_{\mathit{b,a}}}-{{A}_{\mathit{a,b}}}\right) \, \left( {{A}_{\mathit{l,k}}}-{{A}_{\mathit{k,l}}}\right) }{4}.
\]
We can now discard the component definition of $F_{\mu\nu}$ as it has served its purpose, and we would like the final result to appear in terms of $F_{\mu\nu}$ for simplicity.
 Instead, we define the symmetry properties of $F_{\mu\nu}$:
{\color{labelcolor}\begin{verbatim}
 (%i5) remcomps(F)$
 (%i6) decsym(F,0,2,[],[anti(all)])$
 \end{verbatim}}
Next, we compute one of the terms of the Euler-Lagrange equation, and then instruct Maxima to re-express this result in terms of $F_{\mu\nu}$:

{\color{labelcolor}\begin{verbatim}
 (%i7) ishow(diff(L,A([m],[],n)))$
\end{verbatim}}\vskip -1em
\[\color{outputcolor}
 	  -\frac{{{g}^{\mathit{k a}}}\, {{g}^{\mathit{l b}}}\, \left( {{A}_{\mathit{b,a}}}-{{A}_{\mathit{a,b}}}\right) \, \left( {{\mathit{\delta}}_{k}^{n}}\, {{\mathit{\delta}}_{l}^{m}}	-{{\mathit{\delta}}_{k}^{m}}\, {{\mathit{\delta}}_{l}^{n}}\right) }{4}
 	-\frac{{{g}^{\mathit{k a}}}\, {{g}^{\mathit{l b}}}\, \left( {{\mathit{\delta}}_{a}^{n}}\, {{\mathit{\delta}}_{b}^{m}}-
 		{{\mathit{\delta}}_{a}^{m}}\, {{\mathit{\delta}}_{b}^{n}}\right) \, \left( {{A}_{\mathit{l,k}}}-{{A}_{\mathit{k,l}}}\right) }{4}
\]\vskip -1em
{\color{labelcolor}\begin{verbatim}
 (%i8) ishow(subst(F([a,b],[])+A([a],[],b),A([b],[],a),%))$
\end{verbatim}}\vskip -1em
\[\color{outputcolor}
 	  -\frac{{{g}^{\mathit{k a}}}\, {{g}^{\mathit{l b}}}\, {{F}_{\mathit{a b}}}\, \left( {{\mathit{\delta}}_{k}^{n}}\, {{\mathit{\delta}}_{l}^{m}}-{{\mathit{\delta}}_{k}^{m}}\, {{\mathit{\delta}}_{l}^{n}}\right) }{4}-\frac{{{g}^{\mathit{k a}}}\, {{g}^{\mathit{l b}}}\, \left( {{\mathit{\delta}}_{a}^{n}}\, {{\mathit{\delta}}_{b}^{m}}-{{\mathit{\delta}}_{a}^{m}}\, {{\mathit{\delta}}_{b}^{n}}\right) \, \left( {{A}_{\mathit{l,k}}}-{{A}_{\mathit{k,l}}}\right) }{4}
\]\vskip -1em
{\color{labelcolor}\begin{verbatim}
 (%i9) ishow(subst(F([k,l],[])+A([k],[],l),A([l],[],k),%))$
\end{verbatim}}\vskip -1em
\[\color{outputcolor}
 	   -\frac{{{g}^{\mathit{k a}}}\, {{g}^{\mathit{l b}}}\, {{F}_{\mathit{a b}}}\, \left( {{\mathit{\delta}}_{k}^{n}}\, {{\mathit{\delta}}_{l}^{m}}-{{\mathit{\delta}}_{k}^{m}}\, {{\mathit{\delta}}_{l}^{n}}\right) }{4}-\frac{{{g}^{\mathit{k a}}}\, {{g}^{\mathit{l b}}}\, \left( {{\mathit{\delta}}_{a}^{n}}\, {{\mathit{\delta}}_{b}^{m}}-{{\mathit{\delta}}_{a}^{m}}\, {{\mathit{\delta}}_{b}^{n}}\right) \, {{F}_{\mathit{k l}}}}{4}
\]\vskip -1em
{\color{labelcolor}\begin{verbatim}
 (%i10) ishow(canform(rename(contract(expand(%)))))$
\end{verbatim}}\vskip -1em
\[\color{outputcolor}
-{{F}^{\mathit{m n}}}.
\]
Finally, we complete the Euler-Lagrange equation by adding the remaining term:
{\color{labelcolor}\begin{verbatim}
 (%i11) ishow(contract(diff(L,A([m],[])))-idiff(%,n)=0)$
\end{verbatim}}\vskip -1em
 \[\color{outputcolor}
 {{F}_{\mathit{,n}}^{\mathit{m n}}}+{{j}^{m}}=0.
 \]\vskip -1em
\end{ex}

Even this relatively simple example demonstrates how \symb{itensor} can serve as an effective tool for the field theorist.
Similar derivations can be carried out in curved spacetime, making full use of the metric and covariant differentiation, as well as properly accounting for the volume element $\sqrt{-g}$.

The indicial tensor manipulation capability of \symb{itensor} is directly linked to the component tensor manipulation functionality \symb{ctensor}.
A conversion function can translate indexed \symb{itensor} expressions into expression blocks for \symb{ctensor}.
Thus, it is possible to use the indicial capabilities of \symb{itensor}, as in the example above, to derive a set of field equations, which can then be solved by postulating a specific coordinate system and metric using \symb{ctensor}.

\section{Properties of Clifford algebras in view of sparse implementation strategies}
\label{sec:clifprops}

Clifford algebras provide natural generalizations of complex numbers, dual numbers and double numbers.
We will give a brief exposition  on the properties of Clifford algebras over the reals following Dorst \cite{Dorst2001}.

A Clifford algebra is an associative algebra which is generated by a vector space \textit{V} spanned by the orthonormal basis $\lbrace e_1 \ldots e_n\rbrace$, over a field $\mathbb{K}$ of characteristic different from 2.
The algebra is equipped with an anticommutative product, which is referred to as the \textsl{Clifford product}, and
a scalar unit denoted conventionally by $1$. The fundamental relation satisfied by the orthonormal basis is given by
\begin{equation}
 e_i e_j = - e_j e_i, \ \ i \neq j.
\end{equation}
The algebra also has a zero element, denoted conventionally by $0$.

The square of a vector $v$ in the Clifford algebra $C\ell(V, Q)$ is defined by
\begin{equation}
v\,v := Q (v) \, 1, \ \ \ \forall v \in V,
\end{equation}
where $ Q(v)$ denotes a quadratic form.
There are different conventions for the sign of the form \cite{Porteus2000}.
In our implementation, \symb{clifford}, both conventions are allowed, using a sign switch of the pseudo-norm function
\symb{psnorm}.
The latter function depends on the elementary product simplification rules defined by the signature of the algebra (Fig.~\ref{lst:definitions}).

The unit is usually skipped in notation (assuming implicit conversion between the scalar unit of $\mathbb{K}$ and the vector unit of \textit{V} wherever necessary ) and the square of the vector $v$ is denoted conveniently by $v^2$.
The notation $C\ell_{p,q,r} (\mathbb{K})$ ($p+q+r=n$) is interpreted as the convention that
$p$ elements of the orthonormal basis square to $1$, $q$ elements square to $-1$ and $r$ (degenerate) elements square to $0$.
For a given Clifford algebra $C\ell_{p,q,r} (\mathbb{R})$ over the real numbers the quadratic form is, therefore, given by
\begin{equation}
Q(v):= v_1^2+ \ldots + v_p^{2} - v_{p+1}^{2} - \ldots - v_{p+q}^{2}.
\end{equation}

For the elements for which $Q(v) \neq 0$ the inverse element can be reduced into the canonical form
\begin{equation}
v^{-1} = \dfrac{v}{Q(v)}.
\end{equation}

A multivector $A$ of the Clifford algebra is an element of the $2^n$-dimensional vector  space $ P (V) $ spanned by the power set $\{1,e_1,...,e_n,e_1e_2,e_1e_3,...,e_1e_2...e_n\}$.

The inner (dot) product and the outer, or exterior (wedge) product of two elements $a$ and $b$ are defined by
\begin{align}
a \cdot  b := & \dfrac{a\,b + b\,a}{2},  \\
a \wedge b := & \dfrac{a\,b - b\,a}{2}.
\end{align}

A {\em blade} of grade $k$ is an object having the same number of  basis vectors per term.
Conventionally,  $0$-blades are scalars,  $1$-blades are vectors etc.
A general multivector $A$ can be decomposed by the \textit{grade projection operators} $\langle~\rangle_k$ into a sum of different blades:
\begin{equation}
 A = \sum\limits_{k=0}^n \grpart{ A }_k.
\end{equation}

There are two principal involutions -- the Clifford conjugation
$ A^\star $ and the reversion
$ A^\sim $ having the sign mutations shown in Table~\ref{tb:signs}.

\begin{table}
\caption{\label{tb:signs}Sign mutation table for Clifford algebras $C\ell_{p,q,r}$.}
\begin{tabular}{c|cccc}
~&\multicolumn{4}{c}{$k\bmod 4$}\\
~ & 0 & 1 & 2 & 3 \\
\hline
$A^\sim$  & $+$ & $+$ & $-$ & $-$ \\
$A^\star$ & $+$ & $-$ & $-$ & $+$
\end{tabular}
\end{table}

\subsection{Multi-vector derivatives}
\label{sec:mvd}

The vector derivative of a function is well-defined when the Clifford algebra is non-degenerate ($r \neq 0$), which will be assumed further on.
\begin{defn}
	Consider a vector $x$  given in terms of the basis $\left\lbrace  e_1,   \ldots , e_n \right\rbrace $ as $x=\sum_{i=1}^n x^ke_k$.
	For the basis vectors $e_k$ we can define the reciprocal frame $e^k=e_k^{-1}$.
	We now define the vector derivative \cite{Doran2003} as
\begin{equation}
	\mvectdiff{x} f = \sum_ke^k\frac{\partial}{\partial x^k} f(x) =e^k\partial_k f(x) .
\end{equation}
\end{defn}

The vector derivative can be decomposed into inner (directional) and exterior components
\begin{equation}
\mvectdiff{x} F = \mvectdiff{x} \cdot F + \mvectdiff{x} \wedge F
\end{equation}
in a way similar to the decomposition of the geometric product.

\begin{rem}
	The vector derivative can be readily generalized onto general blades if we allow for the arguments to traverse the bases of a general multivector \textit{v}.
	\begin{equation}
	 	\mvectdiff{v} f (x) = \sum\limits_{v_i  \in P(V)}  {\left( v_i\right) }^{-1}  \lim\limits_{ \epsilon \rightarrow 0} \,  \frac{ f(x+ \epsilon \, {v_i}) - f(x) }{   \epsilon }  ,
	\end{equation}
	where $\epsilon >0$ is a scalar vanishing in limit and the limit procedure is understood in norm.
	This definition can be literally implemented in Maxima if proper simplification rules are given for the multivector inverse.
	Such an approach was undertaken in \symb{clifford}.
\end{rem}

\section{The package \symb{clifford}}
\label{sec:cliff}
In Maxima, the Clifford product is represented by the non-commutative operator symbol ($\cdot$).
At present there are two packages that support Clifford algebras in Maxima.
The package \symb{atensor}  partially implements generalized (tensor) algebras  \cite{M2015, Toth2007}.
The \symb{clifford} package implements more extensive functionality related to Clifford algebras\footnote{The code is distributed under  GNU Lesser General Public License from GitHub (\url{http://dprodanov.github.io/clifford/}).}.
The   package relies extensively on the Maxima simplification functionality, and its features are fully integrated into the Maxima simplifier.
The \symb{clifford} package defines multiple rules for pre- and post-simplification of Clifford products, outer products, scalar products, inverses and powers of Clifford vectors.
Using this functionality, any combination of products can be put into a canonical representation.
The main features of the package are summarized in Table \ref{tab:fnct}.

\begin{table}[h]
	\caption{\label{tab:fnct}Main functions in the \symb{clifford} package.}
	\begin{tabular}{l|l}
		function name & functionality \\
		\hline
		\symb{dotsimpc} (ab) & simplification of dot products \\
		\symb{dotinvsimp} (ab)	&  simplification of inverses \\
		\symb{powsimp}	(ab) &  simplification of exponents \\
		\symb{cliffsimpall} (expr) & full simplification of expressions \\
		\symb{cinv} (ab)  & Clifford inverse of element \\
		\symb{grade} (expr) & grade decomposition of expressions\\
		\symb{creverse} (ab) & Clifford reverse of product \\
		\symb{cinvolve} (expr) & Clifford involution of expressions \\
		\symb{cconjugate} (expr) & Clifford conjugate of expressions \\
		\symb{cnorm}(expr) & squared norm of expression \\
	\end{tabular}
\end{table}

\subsection{Command line simplification in \symb{clifford} }
	\label{sec:commsimp}

In the \symb{clifford} package, the inner product is represented by the operator symbol ``$|$'' and the outer (exterior, or wedge) product by the operator symbol ``\&''.
The package defines several geometric product simplification rules which are integrated in the built-in Maxima simplifier (see also Listing~\ref{lst:definitions}).
For example the sum of the inner and outer products of two elements can be immediately combined into the Clifford product:
 {\color{labelcolor}
\begin{verbatim}
 (%i14) a | b + a & b;
\end{verbatim}\vskip -1em
}
$$ \color{outputcolor}  a\cdot b.$$
In a similar manner, the Jacobi identity can be demonstrated to hold for the outer product:

{\color{labelcolor}
\begin{verbatim}
 (%i15) a & b & c + b & c & a + c & a & b;
\end{verbatim}\vskip -1.0em
}
$$ \color{outputcolor}  0.$$

The \symb{clifford} package assigns built-in simplification rules for products of blades  (the first three declarations in Listing~\ref{lst:definitions}) and powers of  blades (the last two  declarations).

Upon initialization, the user has to declare the symbol that will represent the orthonormal basis, and the number of basis vectors with positive, negative and degenerate norms. For example, the \textit{quaternion algebra} can be defined conveniently in \symb{clifford} by issuing the command
  {\color{labelcolor}
   \begin{verbatim}
 (%i21) clifford(e,0,2);
   \end{verbatim}
}
Once the algebra has been initialized, the quaternion multiplication table can be computed by command
{\color{labelcolor}
\begin{verbatim}
 (%i22) mtable2();
\end{verbatim}}\vskip -1em
  	\[\color{outputcolor}
  			\begin{pmatrix}1 & {e}_{1} & {e}_{2} & {e}_{1} . {e}_{2}\cr {e}_{1} & -1 & {e}_{1} . {e}_{2} & -{e}_{2}\cr {e}_{2} & -{e}_{1} . {e}_{2} & -1 & {e}_{1}\cr {e}_{1} . {e}_{2} & {e}_{2} & -{e}_{1} & -1
  			\end{pmatrix}.
  \]

Maxima supports a regime of command line simplification, which can be overloaded by custom functionality.
The \symb{clifford} package defines several simplifying functions, which can be passed as command line options.
In the same algebra $C\ell_{0,2}$, we have:

\begin{itemize}
\item Simplification of inverses:
{\color{labelcolor}\begin{verbatim}
 (%i31) 1/(1+e[1]),cliffsimpall;
\end{verbatim}}\vskip -1em
\begin{flalign*}
{\color{outputcolor}\frac{1 - {e}_{1}}{2}.}
\end{flalign*}
\item Outer product evaluation:
{\color{labelcolor}\begin{verbatim}
 (%i32) e[1]&e[2];
\end{verbatim}}\vskip -1em
\begin{flalign*}
{\color{outputcolor}{e}_{1} . {e}_{2}.}
\end{flalign*}\vskip -1em
{\color{labelcolor}\begin{verbatim}
 (%i33) (1+e[1])&(1+e[1]);
\end{verbatim}}\vskip -1em
\begin{flalign*}
{\color{outputcolor}0.}
\end{flalign*}
\item Inner product evaluation:
{\color{labelcolor}
\begin{verbatim}
 (%i34) (1+e[1])|(1+e[1]);
\end{verbatim}}\vskip -1em
\begin{flalign*}
{\color{outputcolor}2\,{e}_{1}.}\end{flalign*}
\item Multi-vector inverses

A general inverse of a quaternion can be computed. We define an input expression in the form $cc = a + b\, e_1+c \, e_2+d \, e_1 \cdot e_2$ and compute the output as the last part of the command block:
{\color{labelcolor}
\begin{verbatim}
 (%i35) block(declare([a,b,c,d],scalar),
        cc:a+b*e[1]+c*e[2]+d*e[1].e[2],dd:cinv(cc));
\end{verbatim}}\vskip -1em
\begin{flalign*}\color{outputcolor}
dd= \frac{a-{e}_{1}\,b-{e}_{2}\,c-\left( {e}_{1} \cdot {e}_{2}\right) \,d}{{a}^{2}+{b}^{2}+{c}^{2}+{d}^{2}}.
\end{flalign*}

The result can be verified to yield unity after multiplication with the input, expansion and a rational simplification step.
{\color{labelcolor}
\begin{verbatim}
 (%i36) ev(dd.cc,expand,ratsimp)
\end{verbatim}}\vskip -1em
\begin{flalign*}
{\color{outputcolor}1.}\end{flalign*}
\end{itemize}

\section{Geometric calculus functionality in \symb{clifford}}
\label{sec:calculus}

The \symb{clifford} package also implements symbolical differentiation based on the vector derivative. We shall give a presentation of the  Maxwell equations in the geometric algebra $\mathbb{G}^3=C\ell_{3,0}$. The exposition follows closely the presentation of Chappell et al. \cite{Chappell2014}.
In this presentation we assume $c=\mu_0 =\epsilon_0 =1$.

Maxwell's equations can be represented in a convenient way using the notation of Clifford/geometric algebra by a single equation :
\begin{equation}
\nabla_{t+r} F \equiv \left( \mvectdiff{r} +  \frac{\partial }{ \partial t}\right) F = J \, ,
\end{equation}
 where $F$ is the field object (corresponding to the electromagnetic field tensor) and  $J$ is the generalized (4-dimensional) current
\begin{equation}
J =\rho - \left( {{J}_{x}}\, {{e}_{1}}+{{J}_{y}}\, {{e}_{2}}+{{J}_{z}}\, {{e}_{3}}\right),
\end{equation}
using the component representation.
The electromagnetic field object is represented as sum of a vector and a bivector as
\begin{equation}
F = E + i \, B,
\end{equation}
where $i$ is the pseudoscalar of the algebra.
This field object has the required transformation properties under reflection \cite{Chappell2014}.
In the conventional Heaviside-Gibbs notation, the fields are represented by the   vectors $\vec{E}$ and $\vec{B}$, which change sign under reflection of the coordinate system.
On the other hand, a reflection of the coordinate system  the electric field is inverted whereas the magnetic field is invariant.
Hence, an ambiguity is introduced with the conventional Heaviside vector notation, in that, two quantities with different geometric properties are both represented using the same type of mathematical object.

\subsection{Electrostatic and Magnetostatic problems in \symb{clifford}}
\label{sec:static}
\begin{ex}
	In an electrostatic or magnetostatic setting the Green's function of the system
	\begin{equation}
	\nabla_r G = \delta\left( r\right)
	\end{equation}
	is
	\begin{equation}
	G \left(x,y,z \right)  =\frac{{{e}_{1}}\, x+{{e}_{2}}\, y+{{e}_{3}}\, z}{ 4 \pi \ \sqrt{{\left( {{x}^{2}}+{{y}^{2}}+{{z}^{2}}\right) }^{ 3 }}} \, .
	\end{equation}
	Direct calculation issuing the commands
	{\color{labelcolor}\begin{verbatim}
		(%i71) X:e[1]*x+e[2]*y+e[3]*z$
		(%i72) G:X/sqrt(-cnorm(X))^3$
		(%i73) mvectdiff(G,X)$
		\end{verbatim}}\noindent
	evaluates to  0  for $x,y,z\ne 0$.
	In the last calculation the factor is skipped for simplicity.
	$G \left(x,y,z \right)$ can be derived from the following scalar potential
	\begin{equation}
	V(x,u,z) = -\frac{C}{\sqrt{{{x}^{2}}+{{y}^{2}}+{{z}^{2}}}} \, ,
	\end{equation}
	where $C$ is an arbitrary constant matching the initial or boundary conditions.
	Applying the multivector derivative to $V(x,u,z)$ yields
	{\color{labelcolor}\begin{verbatim}
		(%i74) mvectdiff(-1/sqrt(x^2+y^2+z^2),X);
		\end{verbatim}}
	\[\color{outputcolor}
	\frac{{{e}_{1}}\, x+{{e}_{2}}\, y+{{e}_{3}}\, z}{{{\left( {{x}^{2}}+{{y}^{2}}+{{z}^{2}}\right) }^{\frac{3}{2}}}} \, ,
	\]
	which can be recognized as a scalar multiple of the Green's function.
	If we take for example $C= i$ and apply the vector derivative to the potential, $\nabla_r  i V$, we get
	{\color{labelcolor}\begin{verbatim}
		(%i75) mvectdiff(-%iv/sqrt(x^2+y^2+z^2),X);
		\end{verbatim}
	}
	\[\color{outputcolor}
	\frac{\left( {{e}_{2}}\cdot{{e}_{3}}\right) \, x-\left( {{e}_{1}}\cdot{{e}_{3}}\right) \, y+\left( {{e}_{1}}\cdot{{e}_{2}}\right) \, z}{{{\left( {{x}^{2}}+{{y}^{2}}+{{z}^{2}}\right) }^{\frac{3}{2}}}},
	\]
	which is also a solution.
	Therefore, the most general expression for the Green's function is
	\begin{equation}
	G = C_1 \frac{{{e}_{1}}\, x+{{e}_{2}}\, y+{{e}_{3}}\, z}{{{\left( {{x}^{2}}+{{y}^{2}}+{{z}^{2}}\right) }^{\frac{3}{2}}}}
	+ C_2 \frac{\left( {{e}_{2}}\cdot{{e}_{3}}\right) \, x-\left( {{e}_{1}}\cdot{{e}_{3}}\right) \, y+ \left( {{e}_{1}}\cdot{{e}_{2}}\right) \, z}{{{\left( {{x}^{2}}+{{y}^{2}}+{{z}^{2}}\right) }^{\frac{3}{2}}}},
	\end{equation}
	which can be represented in the form
\begin{equation}
	G = \grpart{ G }_1 + \grpart{ G }_2 = E + i  B,
\end{equation}
	corresponding to an electromagnetic field object.
\end{ex}

\subsection{Electromagnetic field in \symb{clifford}}
\label{sec:EM}
\begin{ex}
The multivector $\vec{F}$ representing the electromagnetic field is defined in the end of the following command sequence, in which the variables $\vec{EE}$ and $\vec{BB}$ respectively represent the electric and the magnetic vector fields:
{\color{labelcolor}\begin{verbatim}
 (%i42) EE:cvect(E, [x,y,z]);
\end{verbatim}}\vskip -1em
\[\color{outputcolor}
 {{e}_{1}}\, {{E}_{x}}+{{e}_{2}}\, {{E}_{y}}+{{e}_{3}}\, {{E}_{z}}
\]\vskip -1.0em
{\color{labelcolor}\begin{verbatim}
 (%i43) BB:cvect(B, [x,y,z]);
\end{verbatim}}\vskip -1em
\[\color{outputcolor}
 {{e}_{1}}\, {{B}_{x}}+{{e}_{2}}\, {{B}_{y}}+{{e}_{3}}\, {{B}_{z}}
\]\vskip -1em
{\color{labelcolor}\begin{verbatim}
 (%i44) F:cliffsimpall(EE + %iv.BB);
\end{verbatim}}\vskip -1em
\[\color{outputcolor}
 	 \left( {{e}_{2}}\cdot{{e}_{3}}\right) \,  \, {{B}_{x}}+{{e}_{1}}\, {{E}_{x}}-\left( {{e}_{1}}\cdot{{e}_{3}}\right) \,  \, {{B}_{y}}+{{e}_{2}}\, {{E}_{y}}+\left( {{e}_{1}}\cdot{{e}_{2}}\right) \,  \, {{B}_{z}}+{{e}_{3}}\, {{E}_{z}} \, .
\]
The variable  \symb{\%iv} in \symb{clifford} denotes the pseudoscalar element of $\mathbb{G}^3$. Simplification revealed that $\vec{F}$ is the sum of a vector and a bivector.

The multivector derivative of $\vec{F}$ can be computed further as follows by the following commands:
{\small\color{labelcolor}\begin{verbatim}
 (%i46) r:e[1]*x+e[2]*y+e[3]*z$
 (%i47) DF:mvectdiff(FF1, r)$
 (%i48) G:grade(DF)$
\end{verbatim}}\noindent
where in the last step, we apply grade decomposition, which reveals a scalar sector:
{\small\color{labelcolor}\begin{verbatim}
 (%i49) G[1];
\end{verbatim}}\vskip -1em
\[\color{outputcolor}
 \frac{\partial {{E}_{x}}}{\partial\,x}\, +\frac{\partial {{E}_{y}}}{\partial\,y}\, +\frac{\partial {{E}_{z}}}{\partial\,z}.
\]
A vector in the Euclidean sector can also be identified there as
{\small\color{labelcolor}\begin{verbatim}
 (%i50) G[2];
\end{verbatim}}\vskip -1em
\[\color{outputcolor}
  	 {{e}_{1}}\,  \, \left( \frac{\partial {{B}_{y}}}{\partial\,z}\, -\frac{\partial {{B}_{z}}}{\partial\,y}\, \right)
 	 +{{e}_{2}}\,  \, \left(\frac{\partial {{B}_{z}}}{\partial\,x}\,  -\frac{\partial {{B}_{x}}}{\partial\,z}\, \right)
 	 +{{e}_{3}}\,  \, \left( \frac{\partial {{B}_{x}} }{\partial\,y}\,  -\frac{\partial {{B}_{y}}}{\partial\,x}\, \right),
\]
having the following antisymmetric multiplication table
\[\color{outputcolor}
 \begin{pmatrix}1 & {{e}_{1}} & {{e}_{2}} & {{e}_{3}}\cr {{e}_{1}} & 1 & {{e}_{1}}\cdot{{e}_{2}} & {{e}_{1}}\cdot{{e}_{3}}\cr {{e}_{2}} & -{{e}_{1}}\cdot{{e}_{2}} & 1 & {{e}_{2}}\cdot{{e}_{3}}\cr {{e}_{3}} & -{{e}_{1}}\cdot{{e}_{3}} & -{{e}_{2}}\cdot{{e}_{3}} & 1\end{pmatrix}.
\]
The bivector segment in the field derivative  can be identified as
{\small\color{labelcolor}\begin{verbatim}
 (%i51) G[3];
\end{verbatim}}\vskip -1em
\[\color{outputcolor}
  \left( {{e}_{1}}\cdot{{e}_{2}}\right) \, \left( \frac{\partial  {{E}_{y}} }{\partial\,x}\, -\frac{\partial {{E}_{x}}}{\partial\,y}\,  \right)
  +\left( {{e}_{1}}\cdot{{e}_{3}}\right) \, \left( \frac{\partial {{E}_{z}} }{\partial\,x}\,  -\frac{\partial  {{E}_{x}}}{\partial\,z}\, \right)
  +\left( {{e}_{2}}\cdot{{e}_{3}}\right) \, \left( \frac{\partial {{E}_{z}}}{\partial\,y}\,  -\frac{\partial {{E}_{y}} }{\partial\,z}\, \right).
\]
This segment is isomorphic to the quaternionic algebra, hence it can be referred to as \textit{quaternionic}.
{\small\color{labelcolor}\begin{verbatim}
 (%i52) mtable1([e[1] . e[2],e[2] . e[3], e[1] . e[3]]);
\end{verbatim}}\vskip -1em
\[\color{outputcolor}
  	 \begin{pmatrix}1 & {{e}_{1}}\cdot{{e}_{2}} & {{e}_{2}}\cdot{{e}_{3}} & {{e}_{1}}\cdot{{e}_{3}}\cr {{e}_{1}}\cdot{{e}_{2}} & -1 & {{e}_{1}}\cdot{{e}_{3}} & -{{e}_{2}}\cdot{{e}_{3}}\cr {{e}_{2}}\cdot{{e}_{3}} & -{{e}_{1}}\cdot{{e}_{3}} & -1 & {{e}_{1}}\cdot{{e}_{2}}\cr {{e}_{1}}\cdot{{e}_{3}} & {{e}_{2}}\cdot{{e}_{3}} & -{{e}_{1}}\cdot{{e}_{2}} & -1\end{pmatrix}.
\]

Finally, the \textit{pseudoscalar} sector can be identified by
{\small\color{labelcolor}\begin{verbatim}
 (%i53) G[4];
\end{verbatim}}\vskip -1em
\[\color{outputcolor}
\left( {{e}_{1}}\cdot{{e}_{2}}\cdot{{e}_{3}}\right) \left( \frac{\partial {{B}_{x}}}{\partial\,x}\, +\frac{\partial {{B}_{y}} }{\partial\,y} \, +\frac{\partial {{B}_{z}}}{\partial\,z}\, \right),
\]
with a multiplication table
{\small\color{labelcolor}\begin{verbatim}
 (%i54) mtable1([%iv]);
\end{verbatim}}\vskip -1em
\[\color{outputcolor}
\begin{pmatrix}1 & {{e}_{1}}\cdot{{e}_{2}}\cdot{{e}_{3}}\cr {{e}_{1}}\cdot{{e}_{2}}\cdot{{e}_{3}} & -1\end{pmatrix}.\]
In accordance with Gauss's law for magnetism, no magnetic monopoles exist, which implies that this segment is identically zero.
\end{ex}

Due to the difficulties of representing Maxwell's equations using quaternions, Heaviside rejected Hamilton's algebraic system and developed a system of vector notations using the scalar (dot) and
cross products, which is the conventional system used today in engineering and physics.
On the other hand, representation of Maxwell's equations in $ \mathbb{G}^3$ reveals the quaternionic symmetries in the vector derivative of the electromagnetic field \cite{Gsponer1993, Vlaenderen2001}, which are hidden in the conventional vector and  tensor notations (e.g, Eq.~(\ref{eq:MaxTens})).

\subsection{Homogeneous d'Alembert equation in \symb{clifford}}
\label{sec:paravect}

Clifford algebras offer a convenient way to combine objects of different grades in the form of inhomogeneous sums.
For example, the sum of a scalar and a 3-vector in $\mathbb{G}^3$ (called a \textit{paravector}) is a well-defined inhomogeneous object that can be used in calculations. The Euler-Lagrange field equations corresponding to the Lagrangian density $\mathcal{L}\left(q,\partial_x q \right) $ involving a field $q$ and its derivatives $ \partial_x   q$ with respect to the coordinates $x$ are derived in \cite{Lasenby1993}, which we replicate in the following example using the \symb{clifford} package.

\begin{ex}
Let $q=A$ be the paravector potential given by
\begin{equation}\label{eq:parapot}
A = {{A}_{t}}+{{e}_{1}}\, {{A}_{x}}+{{e}_{2}} \, {{A}_{y}}+{{e}_{3}}\, {{A}_{z}},
\end{equation}
which can be constructed by the command
{\color{labelcolor}\begin{verbatim}
 (%i84) AA: celem(A,[t,x,y,z])$
\end{verbatim}}
Then the geometric derivative object is given by
\begin{equation}
F = \nabla_{t-r} A,
\end{equation}
which is implemented as
{\color{labelcolor}\begin{verbatim}
 (%i85) F:mvectdiff(AA,t-r)$
\end{verbatim}}\noindent
resulting in scalar, vector and bivector parts:
\begin{equation}
F= \grpart{F}_0 +\grpart{F}_1 + \grpart{F}_2.
\end{equation}

The d'Alembert equation for the paravector potential can be derived from a  purely quadratic Lagrangian composed from the components of the geometric derivative

\begin{equation}
\mathcal{L}_a =  \frac{1}{2} \grpart{F^2 }_0
\end{equation}

{\color{labelcolor}
	\begin{verbatim}
	(%i86)  L:lambda([x],1/2*scalarpart(cliffsimpall(x.x)))(F);
	\end{verbatim}
}
\vskip -1em
{\color{outputcolor}
	\begin{flalign*}
	({{\left( {{{{A}_{t}}}_{t}}\right) }^{2}}+{{\left( {{{{A}_{t}}}_{x}}\right) }^{2}}+{{\left( {{{{A}_{t}}}_{y}}\right) }^{2}}+{{\left( {{{{A}_{t}}}_{z}}\right) }^{2}}-2\, \left( {{{{A}_{t}}}_{x}}\right) \, \left( {{{{A}_{x}}}_{t}}\right) +{{\left( {{{{A}_{x}}}_{t}}\right) }^{2}}
	-2\, \left( {{{{A}_{t}}}_{t}}\right) \, \left( {{{{A}_{x}}}_{x}}\right) \\
	+{{\left( {{{{A}_{x}}}_{x}}\right) }^{2}}
	- {{\left( {{{{A}_{x}}}_{y}}\right) }^{2}}-{{\left( {{{{A}_{x}}}_{z}}\right) }^{2}}-2\, \left( {{{{A}_{t}}}_{y}}\right) \, \left( {{{{A}_{y}}}_{t}}\right) +{{\left( {{{{A}_{y}}}_{t}}\right) }^{2}} \\
	+2\, \left( {{{{A}_{x}}}_{y}}\right) \, \left( {{{{A}_{y}}}_{x}}\right)
	-{{\left( {{{{A}_{y}}}_{x}}\right) }^{2}}-2\, \left( {{{{A}_{t}}}_{t}}\right) \, \left( {{{{A}_{y}}}_{y}}\right) \\ +2\, \left( {{{{A}_{x}}}_{x}}\right) \, \left( {{{{A}_{y}}}_{y}}\right)
	+ {{\left( {{{{A}_{y}}}_{y}}\right) }^{2}}-{{\left( {{{{A}_{y}}}_{z}}\right) }^{2}}-2\, \left( {{{{A}_{t}}}_{z}}\right) \, \left( {{{{A}_{z}}}_{t}}\right) +{{\left( {{{{A}_{z}}}_{t}}\right) }^{2}} \\
	+2\, \left( {{{{A}_{x}}}_{z}}\right) \, \left( {{{{A}_{z}}}_{x}}\right) -{{\left( {{{{A}_{z}}}_{x}}\right) }^{2}}+2\, \left( {{{{A}_{y}}}_{z}}\right) \, \left( {{{{A}_{z}}}_{y}}\right) -{{\left( {{{{A}_{z}}}_{y}}\right) }^{2}}-2\, \left( {{{{A}_{t}}}_{t}}\right) \, \left( {{{{A}_{z}}}_{z}}\right)\\
	+2\, \left( {{{{A}_{x}}}_{x}}\right) \, \left( {{{{A}_{z}}}_{z}}\right)
	+2\, \left( {{{{A}_{y}}}_{y}}\right) \, \left( {{{{A}_{z}}}_{z}}\right) +{{\left( {{{{A}_{z}}}_{z}}\right) }^{2}})/\, 2 \ ,
	\end{flalign*}}
where we can recognize the identity
\begin{equation}
\mathcal{L}_a =  \frac{1}{2} \left(  \grpart{F }_0 ^2 + \grpart{F }_1 ^2 + \grpart{F }_2^2 \right)
\end{equation}

{\color{labelcolor}
\begin{verbatim}
(%i87)  S:scalarpart(F)$
(%i88) 	V:vectorpart(F)$
(%i89) 	Q:grpart(F,2)$
(%i90) 	L-1/2*(S.S+V.V+Q.Q),cliffsimpall;
\end{verbatim}
}
 \vskip -1em
\[\color{outputcolor}
0.
\]

Finally, applying the functional derivative, that is the Euler-Lagrange functional yields:
{\color{labelcolor}
	\begin{verbatim}
	(%i50) EuLagEq2(L, t+r,[AA,dA]);
	\end{verbatim}
}
\vskip -1em
{\color{outputcolor}\begin{flalign*}
&  \ \ \ \left( -{{{{A}_{t}}}_{tt}}+{{{{A}_{t}}}_{xx}}+{{{{A}_{t}}}_{yy}}+{{{{A}_{t}}}_{zz}}\right)
+{{e}_{1}}\, \left(  -{{{{A}_{x}}}_{tt}}+{{{{A}_{x}}}_{xx}}+{{{{A}_{x}}}_{yy}}+{{{{A}_{x}}}_{zz}}\right) \\
&+{{e}_{2}}\, \left(  -{{{{A}_{y}}}_{tt}}+{{{{A}_{y}}}_{xx}}+{{{{A}_{y}}}_{yy}}+{{{{A}_{y}}}_{zz}}\right)
+{{e}_{3}}\, \left(  -{{{{A}_{z}}}_{tt}}+{{{{A}_{z}}}_{xx}}+{{{{A}_{z}}}_{yy}}+{{{{A}_{z}}}_{zz}}\right),
\end{flalign*}}
which can be recognized as the D'Alembertian for the components of the paravector potential:
\begin{equation}
 \nabla_{t+r} \nabla_{t-r} A = 0.
\end{equation}
The Maxima expression can be decomposed in a matrix form as
{\color{labelcolor}
	\begin{verbatim}
	(%i51) 	bdecompose(%);
	\end{verbatim}
}

{\color{outputcolor}
	\begin{flalign*}
	[[[1],& \begin{pmatrix}-{{{{A}_{t}}}_{tt}}+{{{{A}_{t}}}_{xx}}+{{{{A}_{t}}}_{yy}}+{{{{A}_{t}}}_{zz}}\end{pmatrix}],\\
	[[{{e}_{1}},{{e}_{2}},{{e}_{3}}], &\begin{pmatrix} -{{{{A}_{x}}}_{tt}}+{{{{A}_{x}}}_{xx}}+{{{{A}_{x}}}_{yy}}+{{{{A}_{x}}}_{zz}}\cr  -{{{{A}_{y}}}_{tt}}+{{{{A}_{y}}}_{xx}}+{{{{A}_{y}}}_{yy}}+{{{{A}_{y}}}_{zz}}\cr  -{{{{A}_{z}}}_{tt}}+{{{{A}_{z}}}_{xx}}+{{{{A}_{z}}}_{yy}}+{{{{A}_{z}}}_{zz}}\end{pmatrix}],\\
	[[0], & \begin{pmatrix}0\end{pmatrix}],[[0],\begin{pmatrix}0\end{pmatrix}]]
	\end{flalign*}
}
if one wishes to solve for individual components.
\end{ex}

\subsection{Euler-Lagrange treatment of electromagnetism using \symb{clifford}}
\label{sec:EL}

\begin{ex}
Finally, we demonstrate the capabilities of the Clifford package by deriving basic identities and the field equations of electromagnetism from Lagrangian functionals.
We begin with loading and setting up the package (see Listing~\ref{lst:geom} for details), establishing the appropriate Clifford algebra, declaring dependencies and defining the coordinates. Next, we define an 4-vector field \textit{A}.

To demonstrate the gauge invariance of electromagnetism, we add to this field the 4-gradient of an arbitrary twice differentiable scalar field $f$, which shall vanish from all the field equations.
We then define the electromagnetic field tensor $F$, which is the exterior derivative of $A$. We compute the vector derivative of $A$, which shall be used later when computing the Euler-Lagrange equation:
{\color{labelcolor}
\begin{verbatim}
 (%i10) a:celem(A,[t,x,y,z])+mvectdiff(f,t+r);
\end{verbatim}}\vskip -1em
\[\color{outputcolor}
{{f}_{t}}+{{e}_{1}}\,   {{f}_{x}} +{{e}_{2}}\,   {{f}_{y}}  +{{e}_{3}}\,   {{f}_{z}} +{{A}_{t}}+{{e}_{1}}\, {{A}_{x}}+{{e}_{2}}\, {{A}_{y}}+{{e}_{3}}\, {{A}_{z}}
\]\vskip -1em
{\color{labelcolor}\begin{verbatim}
 (%i11) dA:mvectdiff(a,t-r);
\end{verbatim}}\vskip -1em
{\color{outputcolor}\begin{align*}
&f_{tt}-f_{xx}-f_{yy}-f_{zz}+{A_{t}}_{t}+e_{1}\left(-{A_{t}}_{x}+{A_{x}}_{t}\right)-{A_{x}}_{x}+e_{1}\cdot e_{2}{A_{x}}_{y}\\
&{}+e_{1}\cdot e_{3}{A_{x}}_{z}+e_{2}\left(-{A_{t}}_{y}+{A_{y}}_{t}\right)-e_{1}\cdot e_{2} \, {A_{y}}_{x}-{A_{y}}_{y}+e_{2}\cdot e_{3}{A_{y}}_{z}\\
&{}+e_{3} \, \left(-{A_{t}}_{z}+{A_{z}}_{t}\right)-e_{1}\cdot e_{3} \, {A_{z}}_{x}-e_{2}\cdot e_{3}{A_{z}}_{y}-{A_{z}}_{z}
\end{align*}}\vskip -1em
{\color{labelcolor}\begin{verbatim}
 (%i12) F:lambda([x],x-scalarpart(x)-grpart(x,4))(mvectdiff(a,t-r));
\end{verbatim}}\vskip -1em
{\color{outputcolor}\begin{align*}
{{{{A}_{t}}}_{t}}+{{e}_{1}}\, \left( -{{{{A}_{t}}}_{x}}+{{{{A}_{x}}}_{t}}\right) -{{{{A}_{x}}}_{x}}+{{e}_{2}}\, \left( -{{{{A}_{t}}}_{y}}+{{{{A}_{y}}}_{t}}\right) \\
-\left( {{e}_{1}}\mathit{ . }{{e}_{2}}\right) \, \left( -{{{{A}_{x}}}_{y}}+{{{{A}_{y}}}_{x}}\right) -{{{{A}_{y}}}_{y}}+{{e}_{3}}\, \left( -{{{{A}_{t}}}_{z}}+{{{{A}_{z}}}_{t}}\right) \\
-\left( {{e}_{1}}\mathit{ . }{{e}_{3}}\right) \, \left( -{{{{A}_{x}}}_{z}}+{{{{A}_{z}}}_{x}}\right) -\left( {{e}_{2}}\mathit{ . }{{e}_{3}}\right) \, \left( -{{{{A}_{y}}}_{z}}+{{{{A}_{z}}}_{y}}\right) -{{{{A}_{z}}}_{z}}.
\end{align*}}

We can now construct a standard EM Lagrangian, $\mathcal{L}_{EM} = - \frac{1}{2} \left\langle F \, F^\star \right\rangle_0$:
{\color{labelcolor}\begin{verbatim}
 (%i13) L:lambda([x],1/2*scalarpart(cliffsimpall(-x.cconjugate(x))))(F);
\end{verbatim}}\vskip -1em
{\color{outputcolor}\begin{align*}
&({{\left( {{{{A}_{t}}}_{x}}\right) }^{2}}+{{\left( {{{{A}_{t}}}_{y}}\right) }^{2}}+{{\left( {{{{A}_{t}}}_{z}}\right) }^{2}}-2\, \left( {{{{A}_{t}}}_{x}}\right) \, \left( {{{{A}_{x}}}_{t}}\right) +{{\left( {{{{A}_{x}}}_{t}}\right) }^{2}}-{{\left( {{{{A}_{x}}}_{y}}\right) }^{2}}-{{\left( {{{{A}_{x}}}_{z}}\right) }^{2}}\\
&{}-2\, \left( {{{{A}_{t}}}_{y}}\right) \, \left( {{{{A}_{y}}}_{t}}\right) +{{\left( {{{{A}_{y}}}_{t}}\right) }^{2}}+2\, \left( {{{{A}_{x}}}_{y}}\right) \, \left( {{{{A}_{y}}}_{x}}\right) -{{\left( {{{{A}_{y}}}_{x}}\right) }^{2}}-{{\left( {{{{A}_{y}}}_{z}}\right) }^{2}}\\
&{}-2\, \left( {{{{A}_{t}}}_{z}}\right) \, \left( {{{{A}_{z}}}_{t}}\right) +{{\left( {{{{A}_{z}}}_{t}}\right) }^{2}}+2\, \left( {{{{A}_{x}}}_{z}}\right) \, \left( {{{{A}_{z}}}_{x}}\right) -{{\left( {{{{A}_{z}}}_{x}}\right) }^{2}}\\
&{}+2\, \left( {{{{A}_{y}}}_{z}}\right) \, \left( {{{{A}_{z}}}_{y}}\right) -{{\left( {{{{A}_{z}}}_{y}}\right) }^{2}})/\, 2.
\end{align*}}
Next, using the vector part of $A$, we construct the electric and magnetic vector fields $E$ and $B$, as well as the current density $j$ and charge density $q$:
{\color{labelcolor}\begin{verbatim}
 (%i14) b:vectorpart(a);
\end{verbatim}}\vskip -1em
{\color{outputcolor}\begin{align*}
{{e}_{1}}\,  {{f}_{x}}  +{{e}_{2}}\,  {{f}_{y}}  +{{e}_{3}}\,   {{f}_{z}} +{{e}_{1}}\, {{A}_{x}}+{{e}_{2}}\, {{A}_{y}}+{{e}_{3}}\, {{A}_{z}}
\end{align*}}\vskip -1em
{\color{labelcolor}\begin{verbatim}
 (%i15) B:factorby(cliffsimpall(-%iv.vvectdiff(b,r)),
        makelist(asymbol[i],i,1,ndim));
\end{verbatim}}\vskip -1em
\[\color{outputcolor}
{{e}_{3}}\, \left( -{{{{A}_{x}}}_{y}}+{{{{A}_{y}}}_{x}}\right) -{{e}_{2}}\, \left( -{{{{A}_{x}}}_{z}}+{{{{A}_{z}}}_{x}}\right) +{{e}_{1}}\, \left( -{{{{A}_{y}}}_{z}}+{{{{A}_{z}}}_{y}}\right) \mbox{}
\]\vskip -1em
{\color{labelcolor}\begin{verbatim}
 (%i16) E:factorby(diff(b,t)-mvectdiff(scalarpart(a),r),
        makelist(asymbol[i],i,1,ndim));
\end{verbatim}}\vskip -1em
\[\color{outputcolor}
{{e}_{1}}\, \left( -{{{{A}_{t}}}_{x}}+{{{{A}_{x}}}_{t}}\right) +{{e}_{2}}\, \left( -{{{{A}_{t}}}_{y}}+{{{{A}_{y}}}_{t}}\right) +{{e}_{3}}\, \left( -{{{{A}_{t}}}_{z}}+{{{{A}_{z}}}_{t}}\right) \mbox{}
\]\vskip -1em
{\color{labelcolor}\begin{verbatim}
 (%i17) j:factorby(cliffsimpall(diff(E,t)-vvectdiff(%iv.B,r)),
        makelist(asymbol[i],i,1,ndim));
\end{verbatim}}\vskip -1em
{\color{outputcolor}\begin{align*}
&{}{{e}_{1}}\, \left( -{{{{A}_{t}}}_{tx}}+{{{{A}_{x}}}_{tt}}-{{{{A}_{x}}}_{yy}}-{{{{A}_{x}}}_{zz}}+{{{{A}_{y}}}_{xy}}+{{{{A}_{z}}}_{xz}}\right)\\
&{}+{{e}_{2}}\, \left( -{{{{A}_{t}}}_{ty}}+{{{{A}_{x}}}_{xy}}+{{{{A}_{y}}}_{tt}}-{{{{A}_{y}}}_{xx}}-{{{{A}_{y}}}_{zz}}+{{{{A}_{z}}}_{yz}}\right)\\
&{}+{{e}_{3}}\, \left( -{{{{A}_{t}}}_{tz}}+{{{{A}_{x}}}_{xz}}+{{{{A}_{y}}}_{yz}}+{{{{A}_{z}}}_{tt}}-{{{{A}_{z}}}_{xx}}-{{{{A}_{z}}}_{yy}}\right)
\end{align*}}\vskip -1em
{\color{labelcolor}\begin{verbatim}
 (%i18) q:svectdiff(E,r);
\end{verbatim}}\vskip -1em
\[\color{outputcolor}
-{{{{A}_{t}}}_{xx}}-{{{{A}_{t}}}_{yy}}-{{{{A}_{t}}}_{zz}}+{{{{A}_{x}}}_{tx}}+{{{{A}_{y}}}_{ty}}+{{{{A}_{z}}}_{tz}}.
\]
Finally, we can obtain the field equations. First, we verify the conservation of charge:
{\color{labelcolor}\begin{verbatim}
 (%i19) svectdiff(q+j,t-r);
\end{verbatim}}\vskip -1em
\[\color{outputcolor}
0.
\]
Next comes Faraday's law:
{\color{labelcolor}\begin{verbatim}
 (%i20) cliffsimpall(-%iv.vvectdiff(E,r)-diff(B,t));
\end{verbatim}}\vskip -1em
\[\color{outputcolor}
0.
\]
Next, Gauss's law for magnetism:
{\color{labelcolor}\begin{verbatim}
 (%i21) svectdiff(B,r);
\end{verbatim}}\vskip -1em
\[\color{outputcolor}
0.
\]
We now check that the postulated Lagrangian is, in fact, identical to the conventional $\frac{1}{2}(E^2-B^2)$ form:
{\color{labelcolor}\begin{verbatim}
 (%i22) 1/2*(E.E-B.B)-L,expand;
\end{verbatim}}\vskip -1em
\[\color{outputcolor}
0.
\]
The Euler-Lagrange equation is identically satisfied:
{\color{labelcolor}\begin{verbatim}
 (%i23) EL:EuLagEq2(L,t+r,[a,dA]);
\end{verbatim}}\vskip -1em
\[\color{outputcolor}
0.
\]\vskip -1em
{\color{labelcolor}\begin{verbatim}
 (%i24) bdecompose(EL);
\end{verbatim}}\vskip -1em
{\color{outputcolor}\begin{align*}
[[[1],& \begin{pmatrix}{{A}_{t}}_{xx}+{{{{A}_{t}}}_{yy}}+{{{{A}_{t}}}_{zz}}-{{{{A}_{x}}}_{tx}}-{{{{A}_{y}}}_{ty}}-{{{{A}_{z}}}_{tz}}\end{pmatrix}],\\
[[{{e}_{1}},{{e}_{2}},{{e}_{3}}], &\begin{pmatrix}{{A}_{t}}_{tx}-{{{{A}_{x}}}_{tt}}+{{{{A}_{x}}}_{yy}}+{{{{A}_{x}}}_{zz}}-{{{{A}_{y}}}_{xy}}-{{{{A}_{z}}}_{xz}}\cr {{A}_{t}}_{ty}-{{{{A}_{x}}}_{xy}}-{{{{A}_{y}}}_{tt}}+{{{{A}_{y}}}_{xx}}+{{{{A}_{y}}}_{zz}}-{{{{A}_{z}}}_{yz}}\cr {{A}_{t}}_{tz}-{{{{A}_{x}}}_{xz}}-{{{{A}_{y}}}_{yz}}-{{{{A}_{z}}}_{tt}}+{{{{A}_{z}}}_{xx}}+{{{{A}_{z}}}_{yy}}\end{pmatrix}],\\
[[0], & \begin{pmatrix}0\end{pmatrix}],[[0],\begin{pmatrix}0\end{pmatrix}]]
\end{align*}}\vskip -1em
{\color{labelcolor}\begin{verbatim}
 (%i25) EL+q+j;
\end{verbatim}}\vskip -1em
\[\color{outputcolor}
0.
\]

Finally, we modify the Lagrangian, introducing an external current $J_0$. Note that gauge invariance is lost in the presence of the external current, and the Lagrangian must be constructed accordingly:
{\color{labelcolor}\begin{verbatim}
 (%i26) dependsv(J,[t,x,y,z])$
 (%i27) A0:celem(A,[t,x,y,z])$
 (%i28) J0:celem(J,[t,x,y,z])$
 (%i29) EL:EuLagEq2(L+scalarpart(A0.J0),t+r,[a,dA])$
 (%i30) bdecompose(EL);
\end{verbatim}}\vskip -1em
{\color{outputcolor}\begin{align*}
[[[1],& \begin{pmatrix}{J_t+{A}_{t}}_{xx}+{{{{A}_{t}}}_{yy}}+{{{{A}_{t}}}_{zz}}-{{{{A}_{x}}}_{tx}}-{{{{A}_{y}}}_{ty}}-{{{{A}_{z}}}_{tz}}\end{pmatrix}],\\
[[{{e}_{1}},{{e}_{2}},{{e}_{3}}], &\begin{pmatrix}{{A}_{t}}_{tx}+J_x-{{{{A}_{x}}}_{tt}}+{{{{A}_{x}}}_{yy}}+{{{{A}_{x}}}_{zz}}-{{{{A}_{y}}}_{xy}}-{{{{A}_{z}}}_{xz}}\cr {{A}_{t}}_{ty}-{{{{A}_{x}}}_{xy}}+J_y-{{{{A}_{y}}}_{tt}}+{{{{A}_{y}}}_{xx}}+{{{{A}_{y}}}_{zz}}-{{{{A}_{z}}}_{yz}}\cr {{A}_{t}}_{tz}-{{{{A}_{x}}}_{xz}}-{{{{A}_{y}}}_{yz}}+J_z-{{{{A}_{z}}}_{tt}}+{{{{A}_{z}}}_{xx}}+{{{{A}_{z}}}_{yy}}\end{pmatrix}],\\
[[0], & \begin{pmatrix}0\end{pmatrix}],[[0],\begin{pmatrix}0\end{pmatrix}]]
\end{align*}}\vskip -1em
{\color{labelcolor}\begin{verbatim}
 (%i31) EL+q+j-J0,expand;
\end{verbatim}}\vskip -1em
\[\color{outputcolor}
0.
\]

This demonstration shows the power of geometric algebra methods when applied to a field theory of physics, and the utility of the \symb{clifford} and \symb{cliffordan} packages to the theoretical physicist.
\end{ex}

\section{Discussion}
\label{sec:disc}

The package \symb{itensor} and the related package \symb{ctensor} allow for advanced symbolical calculation in field theory and general relativity.
Both packages provide complementary functionality that can be used in translating problems from geometric   to tensor calculus and vice versa.

The elements of a Clifford algebra over an $n$-dimensional space live in the $2^n$-dimensional space of the power-set $P(V)$.
Naive implementations of the algebraic structures, therefore, quickly lose performance \cite{Fontijne2007}. Other implementations can use isomorphic matrices over the reals or the complex numbers, which also have a great amount of redundancy \cite{Dorst2007}.
In contrast, sparse representations, as inherently present in Maxima and Lisp, allow for efficient representation of the elements of the algebra.
Another advantage is the possibility to perform complex symbolic calculations, for example the full Clifford simplification following (multi)vector differentiation.

The \symb{clifford} package allows for implementation of functions following closely the mathematical notation used in Clifford and geometric algebra applications.
This allows for implementation of new algorithms following closely their mathematical representation.
Future developments will also incorporate solutions to variational problems, such as derivation of
Euler-Lagrange equations from field Lagrangians.

\subsection{Comparison with other packages}
\label{sec:comp}

Maxima is a general purpose computer algebra system with capabilities that are similar to the two mainstream commercial CAS: Maple and Mathematica.
Unsurprisingly, third-party contributions that implement Clifford, Grassmann and geometric algebra capabilities exist for both products.

For Maple, we note the work the package \symb{CLIFFORD} \cite{Ablamowicz2002}, in development since the late 1990's.
The \symb{CLIFFORD} package was motivated by work on octonions.
The functionality is limited to 9 dimensions.
 Another package, \symb{BIGEBRA} \cite{Ablamowicz2005}, builds on \symb{CLIFFORD}, with the aim to explore Hopf algebras and provide a useful tool for ``experimental mathematics''.

For Mathematica, the \symb{clifford.m} package \cite{Aragon-Camarasa2015} introduces the concept of Clifford and Grassmann algebras, multivectors, and the geometric product. The blades are represented by tuples of numbers.
Some graphical capabilities are also provided.
The authors express their hope that their package, like our Maxima solution, will find its utility in the hands of physicists.

Our Maxima package is also designed to be a useful tool in the hands of applied mathematicians and physicists. Our package emphasizes simplification, including the ability to treat multivectors as sparse objects (i.e., perform algebraic simplification on expressions containing previously not defined symbols representing multivector objects; see Sec.~\ref{sec:commsimp}.)

\subsection*{Authors' contributions}
\label{sec:contrib}

DP is the author of the \symb{clifford}  and \symb{cliffordan} packages. VTT is the author of the Maxima version of the \symb{atensor} package, as well as principal maintainer of \symb{ctensor} and \symb{itensor} for nearly 15 years. The packages \symb{atensor}, \symb{ctensor} and \symb{itensor} are distributed with Maxima. The \symb{clifford} and \symb{cliffordan} packages are freely available for download from GitHub\footnote{\url{http://dprodanov.github.io/clifford/}}. The code of the latter packages is distributed under GNU Lesser General Public License (LGPL).

\section*{Acknowledgments}
The work has been supported in part by a grant from Research Fund - Flanders (FWO), contract numbers
0880.212.840, VS.097.16N.

\bibliographystyle{plain}
\bibliography{clibib}

\begin{thebibliography}{10}

\bibitem{Ablamowicz2002}
R.~{Ablamowicz} and B.~{Fauser}.
\newblock {Mathematics of CLIFFORD - A Maple package for Clifford and Grassmann
  algebras}.
\newblock {\em ArXiv Mathematical Physics e-prints}, December 2002.

\bibitem{Ablamowicz2005}
R.~{Ab{\l}amowicz} and B.~{Fauser}.
\newblock {Clifford and Gra{\ss}mann Hopf algebras via the BIGEBRA package for
  Maple}.
\newblock {\em Computer Physics Communications}, 170:115--130, August 2005.

\bibitem{Aragon-Camarasa2015}
G.~{Aragon-Camarasa}, G.~{Aragon-Gonzalez}, J.~L. {Aragon}, and M.~A.
  {Rodriguez-Andrade}.
\newblock {Clifford Algebra with Mathematica}.
\newblock In I.~J. Rudas, editor, {\em {Recent Advances in Applied
  Mathematics}}, volume~56 of {\em Mathematics and Computers in Science and
  Engineering Series}, pages 64--73, Budapest, 12--14 Dec 2015. Proceedings of
  AMATH '15, WSEAS Press.

\bibitem{Chappell2014}
J.M. Chappell, S.P. Drake, C.L. Seidel, L.J. Gunn, A.~Iqbal, A.~Allison, and
  D.~Abbott.
\newblock Geometric algebra for electrical and electronic engineers.
\newblock {\em Proceedings of the IEEE}, 102(9):1340--1363, 2014.

\bibitem{Doran2003}
C.~Doran and A.~Lasenby.
\newblock {\em Geometric Algebra for Physicists}.
\newblock Cambridge University Press, 2003.

\bibitem{Dorst2001}
L.~Dorst.
\newblock Honing geometric algebra for its use in the computer sciences.
\newblock In G.~Sommer, editor, {\em Geomteric Computing with Clifford
  Algebras}. Springer, 2001.

\bibitem{Dorst2007}
L.~Dorst, D.~Fontijne, and S.~Mann.
\newblock {\em Geometric Algebra for Computer Science. An Object-oriented
  Approach to Geometry}.
\newblock Elsevier, Amsterdam, 2007.

\bibitem{Fontijne2007}
D.~Fontijne.
\newblock {\em Efficient Implementation of Geometric Algebra}.
\newblock PhD thesis, University of Amsterdam, 2007.

\bibitem{Gsponer1993}
A.~{Gsponer} and J.-P. {Hurni}.
\newblock {The physical heritage of Sir W. R. Hamilton. Presented at the
  Conference: ``The Mathematical Heritage of Sir William Rowan Hamilton''
  commemorating the sesquicentennial of the invention of quaternions. Trinity
  College, Dublin, 17th -- 20th August, 1993}.
\newblock {\em ArXiv}, math-ph(0201058), January 2002.

\bibitem{Lasenby1993}
A.~Lasenby, C.~Doran, and S.~Gull.
\newblock A multivector derivative approach to lagrangian field theory.
\newblock {\em Foundations of Physics}, 23(10):1295--1327, 1993.

\bibitem{M2015}
Maxima Project, http://maxima.sourceforge.net/docs/manual/maxima.html.
\newblock {\em Maxima 5.35.1 Manual}, Dec 2014.

\bibitem{Porteus2000}
I.~Porteus.
\newblock {\em Clifford Algebras and the Classical Groups}.
\newblock Cambridge University Press, Cambridge, 2 edition, 2000.

\bibitem{Toth2007}
V.~Toth.
\newblock Tensor manipulation in {GPL} {Maxima}.
\newblock {\em ArXiv}, cs(0503073), 2007.

\bibitem{Vlaenderen2001}
K.J. van Vlaenderen and A.~Waser.
\newblock Generalisation of classical electrodynamics to admit a scalar field
  and longitudinal waves.
\newblock {\em Hadronic Journal}, pages 609 -- 628, 2001.

\end{thebibliography}

\appendix
\section*{Appendix: Example Code listings}

\begin{listing}\caption{\label{lst:tensor}Tensor calculus.}
{\color{labelcolor}\begin{verbatim}
load(itensor);
imetric(g);
components(F([m,n],[]),A([n],[],m)-A([m],[],n));
L:ishow(-1/4*F([k,l])*F([a,b],[])*g([],[k,a])*g([],[l,b])
+j([k],[])*A([l],[])*g([],[k,l]))$
remcomps(F);
decsym(F,0,2,[],[anti(all)]);
ishow(diff(L,A([m],[],n)))$
ishow(subst(F([a,b],[])+A([a],[],b),A([b],[],a),%))$
ishow(subst(F([k,l],[])+A([k],[],l),A([l],[],k),%))$
ishow(canform(rename(contract(expand(%)))))$
ishow(contract(diff(L,A([m],[])))-idiff(%,n)=0)$
\end{verbatim}}
\end{listing}

\begin{listing}
\caption{\label{lst:geom}Geometric calculus.}
{\color{labelcolor}\begin{verbatim}
/* Initialization */
load("clifford.mac")$
load("cliffordan.mac")$
derivabbrev:true$
%divsimp:true$
clifford(e,3)$
declare([t,x,y,z,f], scalar)$
dependsv(A,[t,x,y,z])$
depends(f,[x,y,z,t])$
r:cvect([x,y,z]);
/* The 4-potential with gauge freedom */
a:celem(A,[t,x,y,z])+mvectdiff(f,t+r);
dA:mvectdiff(a,t-r);
/* The electromagnetic field */
F:lambda([x],x-scalarpart(x)-grpart(x,4))(mvectdiff(a,t-r));
/* The Maxwell Lagrangian */
L:lambda([x],1/2*scalarpart(cliffsimpall(-x.cconjugate(x))))(F);
/* spatial field variable */
b:vectorpart(a);
/* The magnetic field */
B:factorby(cliffsimpall(-%iv.vvectdiff(b,r)),makelist(asymbol[i],i,1,ndim));
/* The electric field */
E:factorby(diff(b,t)-mvectdiff(scalarpart(a),r),makelist(asymbol[i],i,1,ndim));
/* The current density and charge density, which are conserved */
j:factorby(cliffsimpall(diff(E,t)-vvectdiff(%iv.B,r)),makelist(asymbol[i],i,1,ndim));
q:svectdiff(E,r);
svectdiff(q+j,t-r);
/* Maxwell's equations */
cliffsimpall(-%iv.vvectdiff(E,r)-diff(B,t));
svectdiff(B,r);
/* Verifying the Lagrangian */
1/2*(E.E-B.B)-L,expand;
/* The Euler-Lagrange equation */
EL:EuLagEq2(L,t+r,[a,dA]);
bdecompose(EL);
EL+(q+j);
/* Introducing an external current */
dependsv(J,[t,x,y,z]);
A0:celem(A,[t,x,y,z]);
J0:celem(J,[t,x,y,z]);
EL:EuLagEq2(L+scalarpart(A0.J0),t+r,[a,dA]);
bdecompose(EL);
EL+q+j-J0,expand;
\end{verbatim}}
\end{listing}

\clearpage

\end{document}